\documentclass[a4paper,11pt]{article}
\usepackage{pos}
\usepackage{amsmath}
\usepackage{booktabs}

\title{Search for high-energy neutrino sources from the direction of IceCube alert events}
 \ShortTitle{IceCube alert followup}

\author{The IceCube Collaboration \\{\normalsize \normalfont(a complete list of authors can be found at the end of the proceedings)}}

\emailAdd{martina.karl@icecube.wisc.edu}

\abstract{We use IceCube's high-statistics, neutrino-induced, through-going muon samples to search for astrophysical neutrino sources. Specifically, we analyze the arrival directions of IceCube's highest energy neutrinos. These high-energy events allow for a good angular reconstruction of their origin. Additionally, they have a high probability to come from an astrophysical source. On average, 8 neutrino events that satisfy these selection criteria are detected per year. Using these neutrino events as a source catalog, we present a search for the production sites of cosmic neutrinos. In this contribution we explore a time-dependent analysis, and present preliminary 3$\sigma$ discovery potential fluences of $\approx 2.7 \cdot 10^{-2} \rm{GeV}/\rm{cm}^2$. We construct the fluences using expectation maximization.\\

\vspace{4mm}
{\bfseries Corresponding authors:}
Martina Karl$^{1, 2 *}$, Philipp Eller$^{2}$, Anna Schubert$^{2}$\\
{$^{1}$ \itshape Max Planck Institute for Physics, F\"ohringer Ring 6, 80805 M\"unchen, Germany}\\
{$^{2}$ \itshape Technical University of Munich, James-Frank-Str. 1, 85748 Garching, Germany}\\[4mm]
$^*$ Presenter

\FullConference{37$^{\rm{th}}$ International Cosmic Ray Conference (ICRC 2021)\\
		July 12th -- 23rd, 2021\\
		Online -- Berlin, Germany}

}

\begin{document}
\maketitle

\section{Introduction}\label{sec:info}

\begin{figure}
    \centering
    \includegraphics[width=0.7\textwidth]{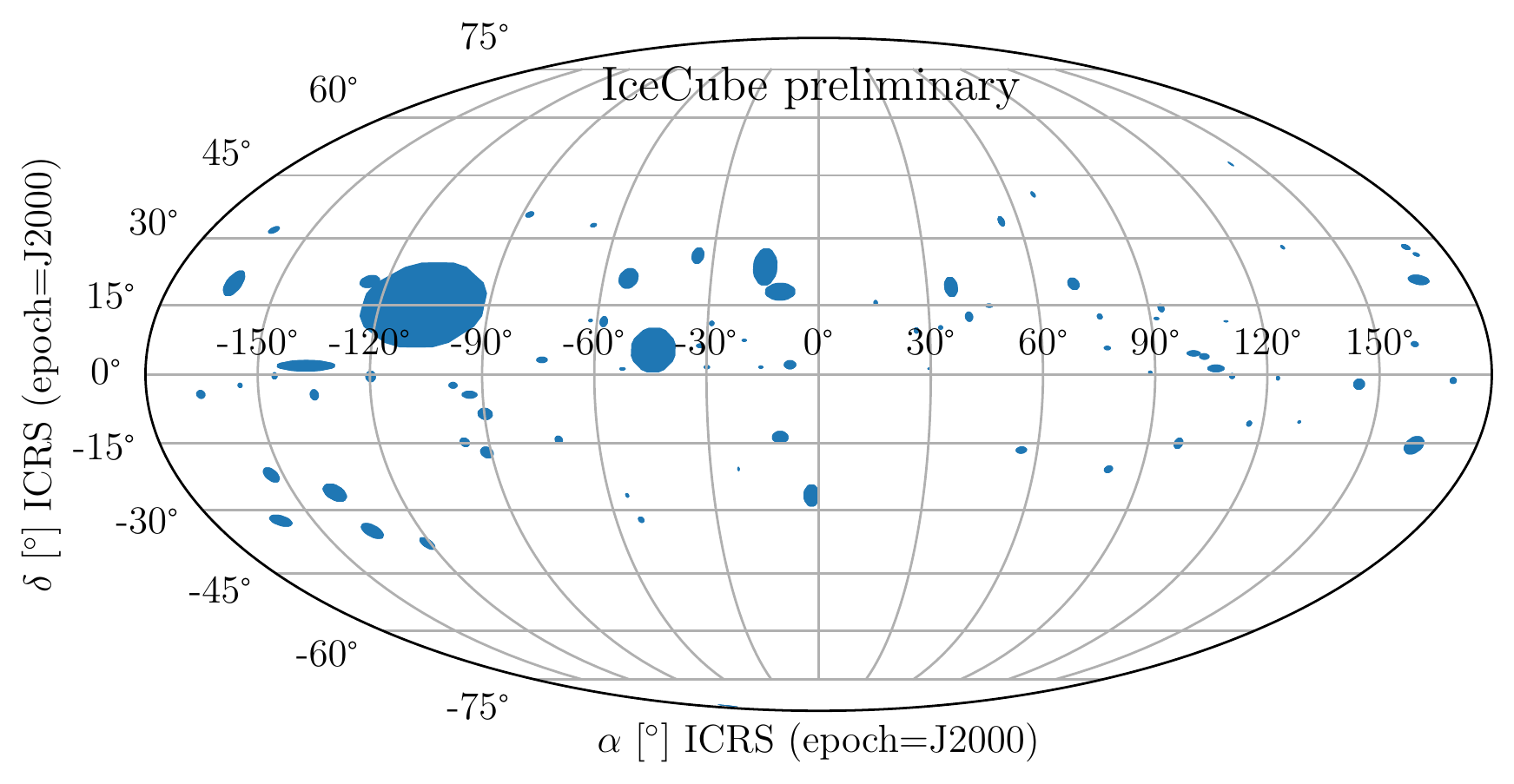}
    \caption{Sky map with the arrival directions of events fulfilling the IceCube alert criteria from August 2009 until March 2019. The size of a dot represents the uncertainty region of the reconstruction, providing a 90\% chance of the alert to be from inside the filled region. Figure taken from \cite{2019ICRC...36..929K}.}
    \label{fig:skymap}
\end{figure}

The IceCube Neutrino Observatory \cite{Aartsen_2017} is  a  cubic-kilometer  scale  neutrino  detector  instrumenting a gigaton of ice at the geographic South Pole in Antarctica. 
Contrary to traditional telescopes, IceCube's field of view comprises the whole sky with the greatest sensitivity for high-energy events at the horizon. It is thus ideally suited to inform other telescopes of interesting events. If a neutrino event has a high probability to be of astrophysical origin, IceCube sends alerts to other telescopes \cite{2017APh....92...30A}. These notifications trigger follow-up multi-messenger observations \cite{Kintscher_2016}. A map of the arrival directions of IceCube alerts is shown in fig. \ref{fig:skymap}. This map shows all events that fulfill the alert criteria \cite{2017APh....92...30A}, starting from August 2009, until March 2019. 

On the 22nd of September 2017, IceCube detected an astrophysical neutrino (IceCube-170922A) with an extremely high energy (EHE alert). The promptly triggered gamma-ray follow-up observations detected a flaring blazar at the origin of this event \cite{2018Sci...361.1378I}. Additionally, we searched for previous neutrino emission from the origin direction in archival IceCube data. We identified a neutrino flare from the same direction between September 2014 and March 2015 \cite{2018Sci...361..147I}. 

This discovery leads to the question whether there are additional neutrino emissions coming from the origins of other alert events. To address this, we analyze 12 years of archival IceCube data and search for an excess in neutrino emission. The IceCube alerts provide positions of interest, comparable to a catalog of possible neutrino sources. A search for steady neutrino sources at the position of alert events was presented in \cite{2019ICRC...36..929K}. In this specific analysis, we search for transient neutrino sources.

\section{Analysis Method}\label{sec:refs}
We expand the time integrated analysis method (presented in \cite{2019ICRC...36..929K}) with a time dependency. We use an unbinned likelihood approach.

\subsection{Point source search with unbinned likelihood ratio}
We expect two components in the neutrino-induced muon sample. One is the astrophysical signal, the other component is the atmospheric background \cite{Braun:2008bg}. Thus, the likelihood is a superposition of the signal ($S$) and background ($B$) probability density functions (pdfs). It is constructed by the product over all events $i$ in the sample. We can neglect the Poisson dependence due to the high number of events in the sample:

\begin{equation}
\mathcal{L} = \prod_{i} \left[ \frac{n_s}{N} S_i (\vec{x}_i , \sigma_i, E_i ; \vec{x}_s, \gamma, t_i, t_s) + \left( 1 - \frac{n_s}{N} \right) B_i (\delta _i, E_i, livetime) \right] .
\end{equation}
Here, $n_s$ denotes the mean number of expected signal events in the detector and is divided by $N$, which is the total number of all detected events (background + signal). The reconstructed origin position for each event $i$ is $\vec{x}_i = (\alpha_i, \delta_i)$, the right ascension $\alpha$ and the declination $\delta$. The one sigma uncertainty of this reconstructed position is given with $\sigma _ i$. The event's energy is given as $E_i$, and the time the event was detected is given as $t_i$. We also consider the source properties: the source position $\vec{x}_s$, and the source energy spectral index $\gamma$. We assume a power law emission spectrum of $\propto E^{-\gamma}$. The source has a time dependency, which is denoted as $t_s$ in the form of flaring time and flare duration. The total detector up-time of the considered data period is called $livetime$. The signal pdf $S_i$ can be split into different parts $S(\vec{x}_i , \sigma_i, E_i ; \vec{x}_s, \gamma, t_i, t_s) = S_{spatial} \cdot S_{energy} \cdot S_{temporal}$ \cite{Braun:2008bg}:

\begin{equation}\label{eq:signal_spatial_signal_energy}
S_{spatial} \cdot S_{energy} = S_i (\vec{x}_i, \sigma_i ; \vec{x}_s) \cdot \varepsilon_{s} (E_i ; \delta _i , \gamma ) = \frac{1}{2 \pi \sigma_i ^2} \exp \left( - \frac{ | \vec{x}_i - \vec{x}_s | ^2}{2 \sigma _i ^2} \right) \cdot \varepsilon _{s} (E_i ; \delta _i, \gamma ),
\end{equation}

\begin{equation}\label{eq:sig_pdf_temporal_gauss}
S_{temporal}(t_i, \mu_T, \sigma_T) = \frac{1}{{\sigma_T \sqrt {2\pi } }}e^{{{ - \left( {t_i - \mu_T } \right)^2 } \mathord{\left/ {\vphantom {{ - \left( {t_i - \mu_T } \right)^2 } {2\sigma_T ^2 }}} \right. \kern-\nulldelimiterspace} {2\sigma_T ^2 }}}.
\end{equation}
The spatial part shows a Gaussian distribution with the source position as mean and the event reconstruction uncertainty as standard deviation. We take the uncertainty of the source position into account in section \ref{sec:position}. The energy factor $\varepsilon _{s}$ is the probability density function for a signal event of energy $E_i$ depending on its declination $\delta _i$, and the source spectral index $\gamma$. This factor is calculated from Monte Carlo simulation \cite{Braun:2008bg}. In the time pdf, we assume that neutrino flares are emitted in a Gaussian shape with the mean $\mu_T$ and the standard deviation $\sigma_T$.

Similarly, we express the background pdf $B_i$ as  
$B(\vec{x}_i, E_i, livetime) =B_{spatial} \cdot B_{energy} \cdot B_{temporal}$ \cite{Braun:2008bg}:

\begin{equation}\label{eq:bg_pdf}
B \left( \vec{x}_i, E_i, livetime\right) = B_i (\vec{x}_i) \cdot \varepsilon _{B} (E_i; \delta _i) \cdot B(livetime) = \frac{1}{2 \pi} \cdot P(\delta _i) \cdot \varepsilon _{B}(E_i; \delta _i) \cdot \frac{1}{livetime},
\end{equation}
because of Icecube's unique geographic location and symmetry, we can assume uniformity over right ascension for background events.\footnote{The IceCube detector is at the Geographic South Pole. Due to earth rotation, we see the same uniform background from all directions in right ascension for integration times of longer than a day.} 
The spatial background pdf depends only on declination $\delta _i$, because of detector geometry. Here, similar to the signal case, the energy term $\varepsilon _{B}$ denotes the probability density function for a background event with energy $E_i$ at declination $\delta _i$. In the temporal background we assume a uniform distribution over the whole detector up-time of IceCube ($livetime$). 

We take the likelihood ratio of the null-hypothesis (background only, $n_s = 0$) and the best fit of the signal hypothesis $n_s > 0$ to calculate the test statistics (TS). We also need to consider the look-elsewhere effect, since there are more short time flare windows than long time flare windows. We correct this by introducing a penalty factor ($pf$) \cite{Braun:2008bg}. The penalty factor depends on the maximum allowed flare length, which is 300~days. The test statistics can be expressed as follows \cite{Braun:2008bg}: 


\begin{equation}\label{eq:TS}
\begin{split}
     TS = 2\left( \log \left[\frac{P(Data|H_S)}{P(Data|H_0)} \right] - \log \left(pf\right) \right) 
    &= 2 ~\left( \sum_{i} \log \left[\frac{\hat{n}_s}{N_{obs}}\left(\frac{S_i}{B_i}-1\right) +1\right]  - \log \left( \frac{300~\rm{days}}{\sqrt{2 \pi} \sigma_T} \right) \right).
\end{split}
\end{equation}

\subsection{Position fit}\label{sec:position}
\begin{figure}
    \centering
    \begin{minipage}{0.49\textwidth}
        \includegraphics[width=\textwidth]{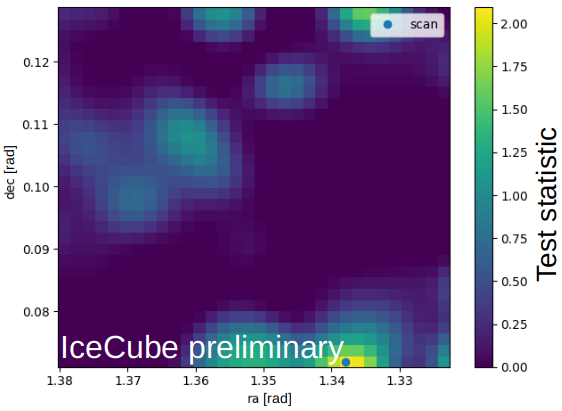}
    \end{minipage}
    \begin{minipage}{0.49\textwidth}
        \includegraphics[width=\textwidth]{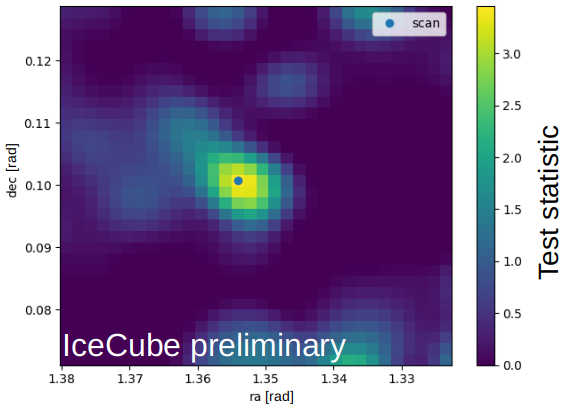}
    \end{minipage}
    \caption{We show example plots of a reconstructed alert 90\% uncertainty region, the test statistic values for a random background distribution (left) and a region with injected signal (right). We divide each region into steps with $0.1 ^{\circ}$ spacing. We determine the best test statistic value at each step by fitting the best mean number of signal events and energy spectral index. We consider the position with the highest test statistic value as the point source position, indicated with the blue dot ("scan"). In the left plot the scan finds the position with the most significant background fluctuation. In the right plot, we simulate 10 signal events injected at the center of the region. The scan returns the position with the simulated signal. Figure taken from \cite{2019ICRC...36..929K}.}
    \label{fig:pos_scan_example}
\end{figure}

We consider alert events in IceCube to be a source catalog for potential high-energy neutrino sources. The alerts do not provide a precise, point-like position of sources. The 90\% error regions are shown in fig. \ref{fig:skymap}. 
We are looking for point-like sources within the 90\% reconstructed uncertainty region of alerts. For this, we divide the uncertainty region in an evenly spaced grid of $0.1 ^{\circ}$ in right ascension and declination. This grid size is smaller than the reconstruction angular resolution, thus this method represents an unbinned search. At each point, we fit the maximal test statistic value depending on the mean number of signal events and the source energy spectral index. Eventually, we select the point yielding the highest test statistic value as the source position. This approach was also used in \cite{2019ICRC...36..929K}. We illustrate this procedure in figure \ref{fig:pos_scan_example}. With this best test statistic value, we build the test statistic distribution and determine a p-value for each alert.

\subsection{Search for neutrino flares}
At each point in the position grid, we search the data for possible time-dependent neutrino flares. Previous approaches followed a brute force testing of different neutrino flares \cite{2018Sci...361..147I}. If we combine the brute force flare search with the brute force position search (see previous paragraph \ref{sec:position}), we need prohibitively large computational resources for calculation of the test statistic. Thus, we apply a new method for the search for neutrino flares in IceCube data: we use expectation maximization \cite{10.2307/2984875} (EM), an unsupervised learning algorithm. This new approach speeds up the analysis significantly (about a factor of $10000$). An in-depth explanation of the expectation maximization algorithm can be found in \cite{press2007numerical}. 

\begin{figure}
    \centering
    \begin{minipage}{0.5\textwidth}
        \includegraphics[width=\textwidth]{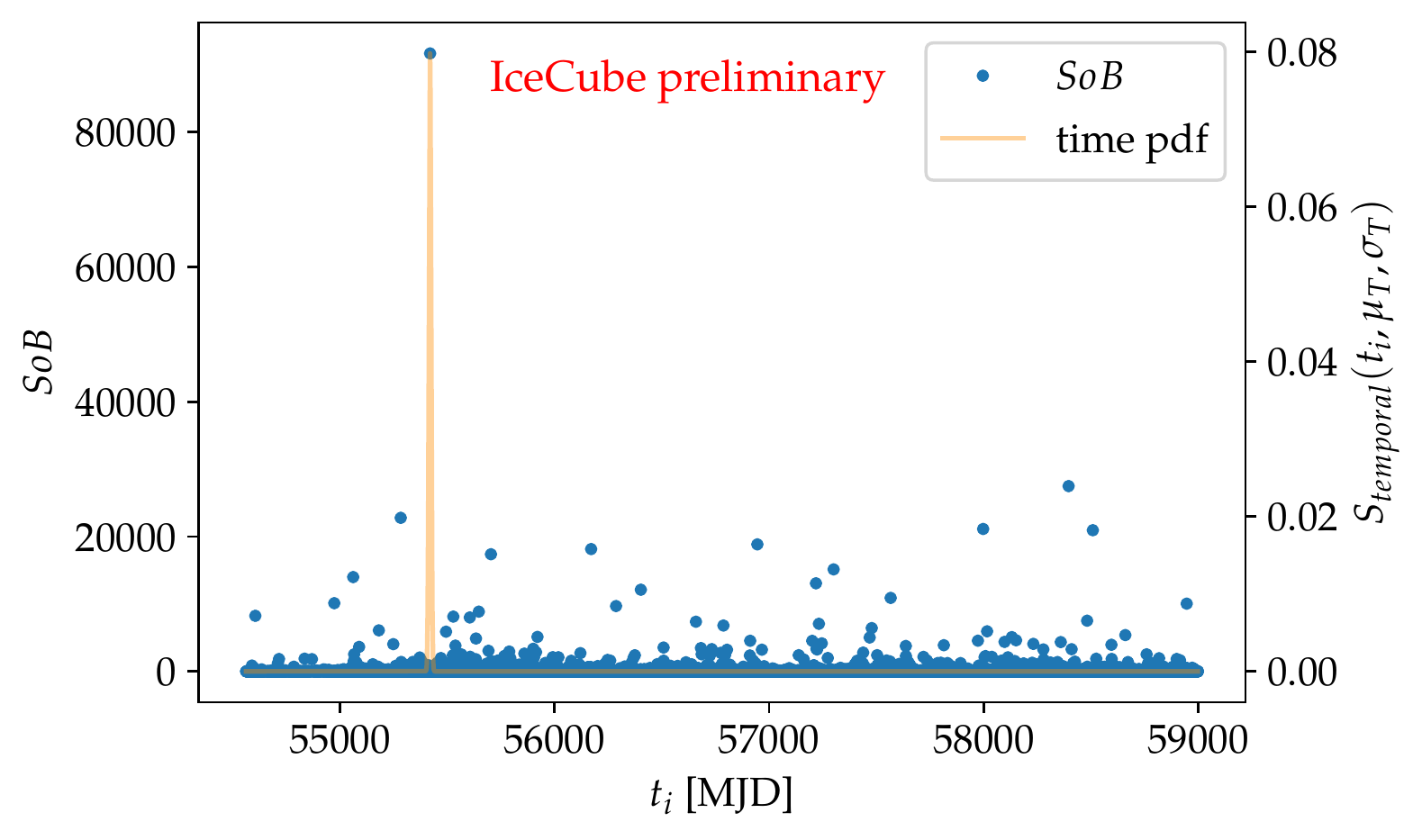}
    \end{minipage}
    \begin{minipage}{0.49\textwidth}
    \centering
     \raisebox{1mm}{\includegraphics[width=0.98\textwidth]{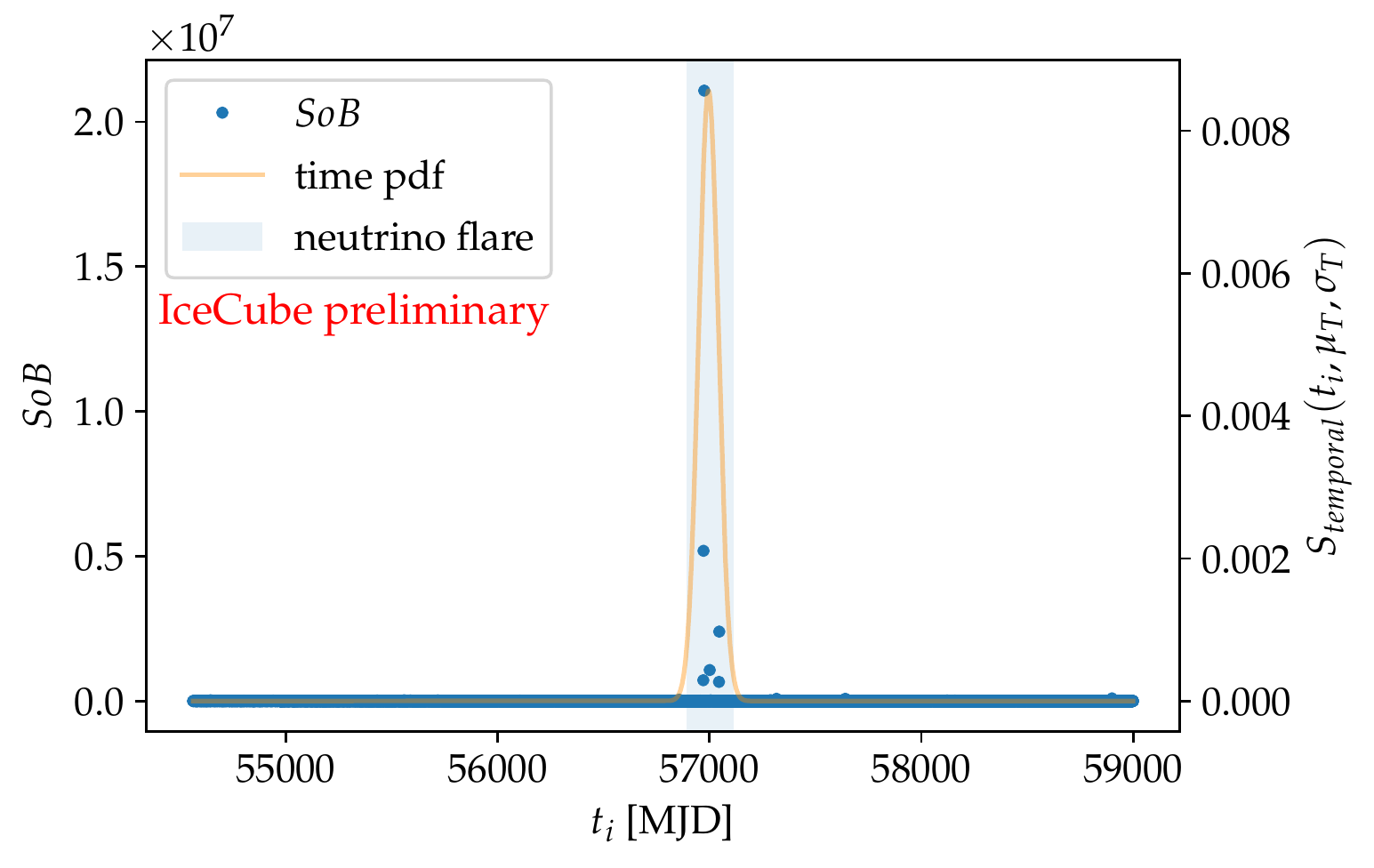}}

    \end{minipage}
    \caption{Left: The $SoB$ ratios for background. The EM algorithm fits background fluctuations (orange curve).\\
    Right: We simulate a neutrino flare of 10 events within 110~days (blue shaded region), the $SoB$ values of these events exceed the background $SoB$ by some orders of magnitude. The EM algorithm finds the flare and determines proper parameters ($\mu_T, \sigma_T$) for the time pdf (orange Gaussian curve).}
    \label{fig:em_example}
\end{figure}

We use a mixture model, with a Gaussian signal pdf (the neutrino flare in time) and a uniform background distribution. We take the spatial and energy information as our data and want to determine the flaring time. We use the spatial and energy signal pdf over background pdf ratio 
$SoB = \frac{S_{spatial} \cdot S_{energy}}{B_{spatial} \cdot B_{energy}}$,
with each pdf defined as in equations \ref{eq:signal_spatial_signal_energy} and \ref{eq:bg_pdf}. Then we apply the EM algorithm with the mixture model on the $SoB$ over time distribution. Hence, we find the best fitting Gaussian time pdf for events that are close to the source and could follow the source spectral distribution. We show an example of the EM algorithm in fig. \ref{fig:em_example}. Left we see the $SoB$ distribution vs. time for background data only and the best fit of background fluctuations. On the right we simulated a neutrino flare of 10 events within 110~days. Their $SoB$ values exceed the background $SoB$ by some orders of magnitude because of their spatial clustering and higher energies. The EM algorithm then maximizes the likelihood and determines the best flaring time. We can then calculate the temporal pdf and determine the test statistic value as in eq. \ref{eq:TS}. We repeat this procedure at every grid point in the uncertainty area (see section \ref{sec:position}).

\subsection{Parametrized description of the test statistic quantiles depending on flare parameter}\label{sec:TS_analytical_description}

The search for a clustering in time in addition to a clustering in space means highly increased computational effort. We investigate an analytical description of how the test statistic distribution changes for different flare intensities.

In this test, we specifically want to investigate the effect of the flare itself. We eliminate other influences by assuming we know the correct position and the correct flare parameters. We also consider a box-shaped neutrino flare model. The box-shaped pdf allows only neutrino emission between the flare starting time $t_{start}$ and the flare end time $t_{end}$. The temporal signal pdf is thus 0 for $t_i \notin [t_{start}, t_{end}]$ and $\frac{1}{\Delta t}$ for $t_i \in [t_{start}, t_{end}$], where $\Delta t = t_{end} - t_{start}$. 


We use this time pdf in equation \ref{eq:TS}. We find the dependency $TS \propto \log \left( \frac{n_s}{\Delta t} \right)$ and define the fitting function:

\begin{equation}\label{eq:TS_fit}
    TS \propto a + b \cdot \text{log} \left( \frac{n_s}{\Delta t} \right).
\end{equation}

\begin{figure}
    \centering
    \includegraphics[width=0.6\textwidth]{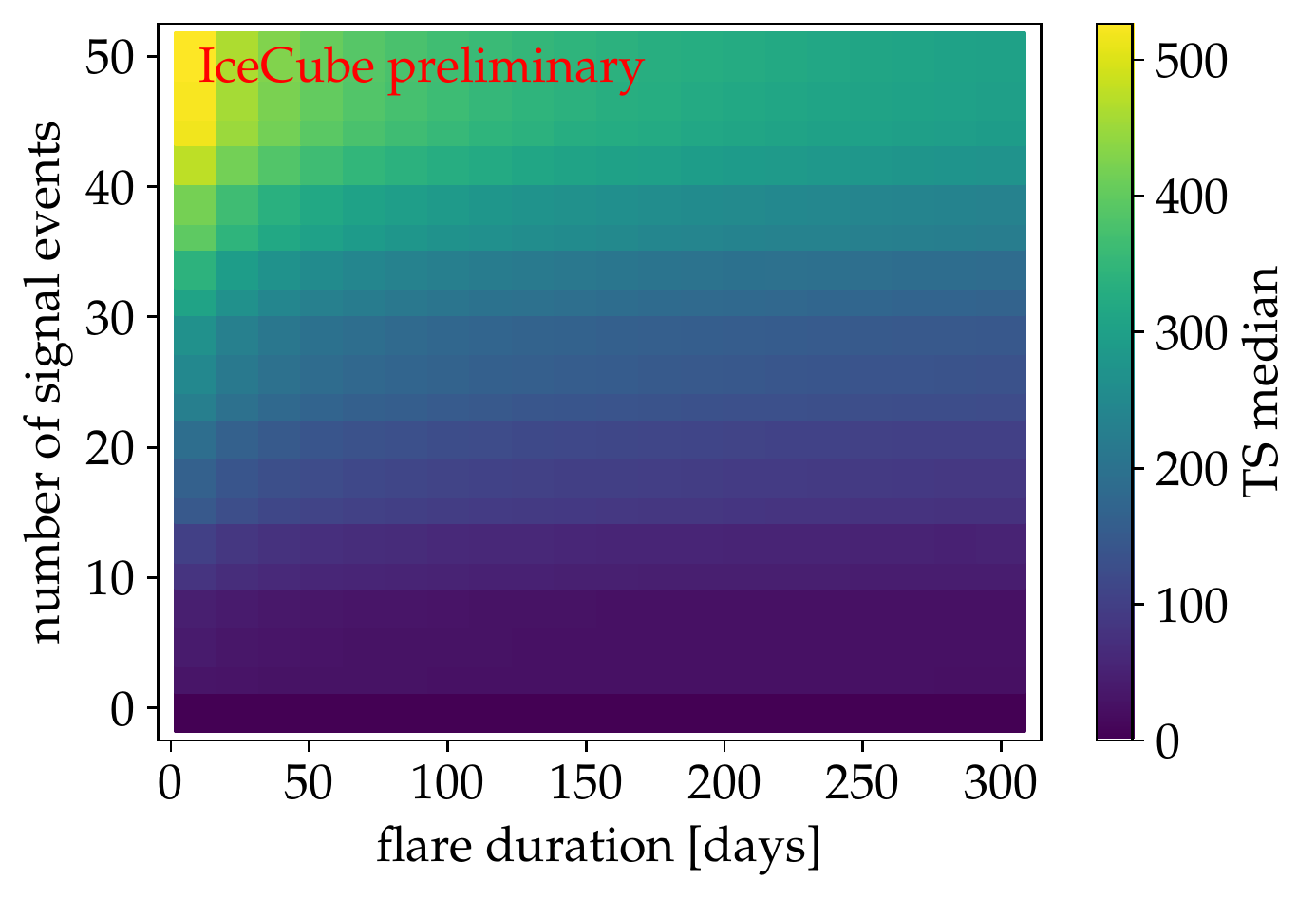}
    \caption{The test statistic median for different flares. We simulated flares with a duration of up to 300~days (x-axis) and with a signal strengths of up to 50 neutrinos (y-axis). Short strong flares yield very high test statistic values, which decrease with increasing flaring time or weaker flare strength.}
    \label{fig:grid}
\end{figure}

\begin{figure}
\centering
\begin{minipage}{.49\textwidth}
  \centering
  \includegraphics[width=\textwidth]{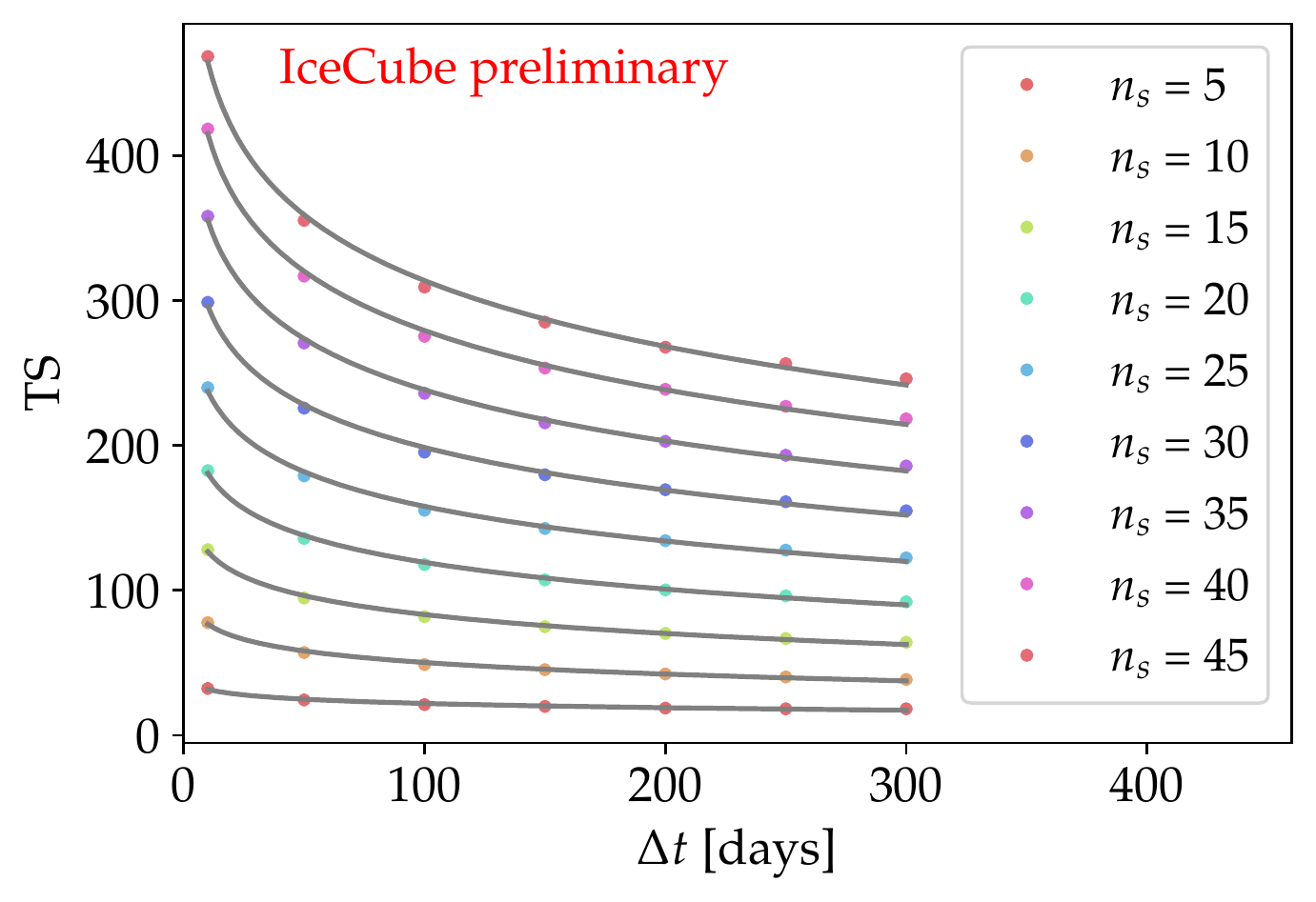}
\end{minipage}
\begin{minipage}{.5\textwidth}
  \centering
  \includegraphics[width=\textwidth]{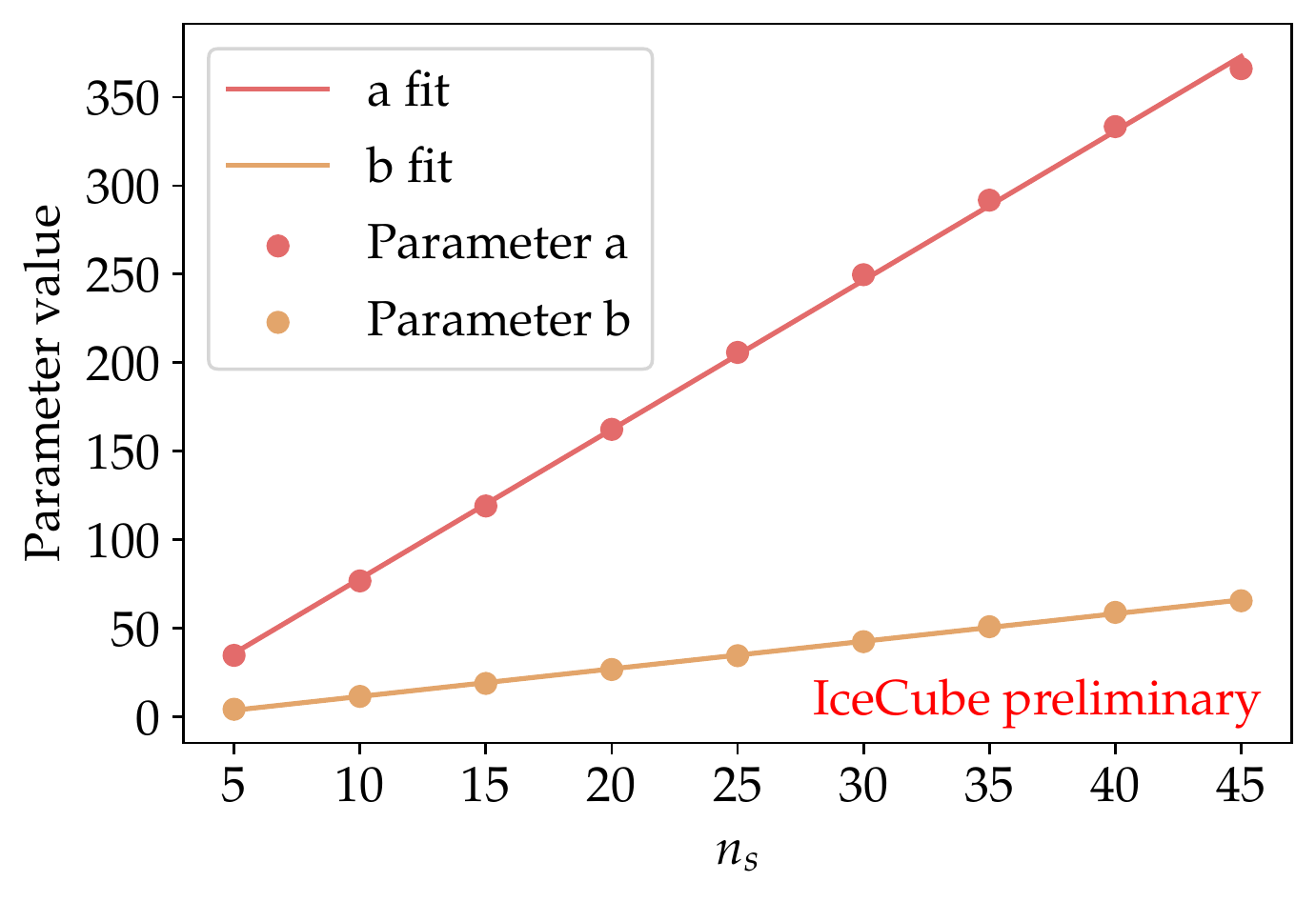}
\end{minipage}%
\caption{Left: the points show the simulated test statistic median for different fixed number of signal events (ninj). The black line is the fit with the function $TS = a + b \cdot \text{log} \left( \frac{n}{\Delta t} \right)$. For each flare strength, the fit accurately describes how the test statistic median changes with increasing flare duration. \\
Right: We take the parameters $a$ and $b$ of the fitting function $TS = a + b \cdot \text{log} \left( \frac{n}{\Delta t} \right)$. The fit paramaters show a linear dependency on the flare strength.}
\label{fig:fit}
\end{figure}

We simulate different flare strengths and flare durations and calculate the test statistic median for each flare (see fig. \ref{fig:grid}). For a fixed flare strength, we fit the test statistic median with the logarithmic function in eq. \ref{eq:TS_fit}, shown in the left of fig. \ref{fig:fit}. The parameters $a$ and $b$ show a linear dependency on the flare strength (see right of fig. \ref{fig:fit}). With this, we can analytically sample the test statistic quantiles for different flares.

\section{Performance study}
We examine the performance of the analysis. A measure for this is the discovery potential. The $3\sigma$ discovery potential is defined as the source flux for which we have a 50\% chance to determine a p-value of at least $3 \sigma$.

In the time-dependent case, a relevant quantity is the total neutrino emission seen from the source during the neutrino flare -- the fluence. The fluence is the source flux integrated over flaring time. We present the  $3 \sigma$ discovery potential fluences of this analysis at the position of the blazar TXS0506+056 ($\alpha = 77.366^{\circ}, \delta = 5.69^{\circ}$) and for the same flare duration as discovered in \cite{2018Sci...361..147I} ($\sigma_T = 55.18~\rm{days}, \Delta t = 110.36~\rm{days}$). We assume a source spectral index of $\gamma = 2$.

We combine several data sets from different data taking periods of IceCube. We want to determine the overall $3\sigma$ discovery potential fluence. Additionally, we want to find out whether these different data sets influence which source flux we can potentially detect. Thus we simulate similar neutrino flares in each data sample and evaluate the fluence of the respective $3\sigma$ discovery potential. We also compare how the shape of the time pdf influences the $3\sigma$ discovery potential fluence. For this, we additionally simulate a neutrino flare following a box-shaped time pdf (as described in section \ref{sec:TS_analytical_description}). 

In table \ref{tab:three-sigma-discovery-potential}, we show the $3\sigma$ discovery potential fluence for each data taking period. We see a slight difference in fluences for different data samples. We calculate the $3\sigma$ discovery potential fluence for a box-shaped time pdf for the latest and longest data set. The fluence for the box-shaped pdf agrees with the fluence of the Gaussian time pdf (see the last two lines of table \ref{tab:three-sigma-discovery-potential}). The mean $3\sigma$ discovery potential fluence is $ 2.7 \cdot 10^{-2} \rm{GeV}/\rm{cm}^2$.

\begin{table}[]
    \centering
    \begin{tabular}{ccc}
        \toprule
         Time pdf shape &Duration of data taking period [days] &  $3\sigma$ discovery potential fluence [$\frac{\rm{GeV}}{\rm{cm}^2}$] \\
         \midrule
         Gaussian &409 & 0.027\\
         \midrule
         Gaussian &376 & 0.037\\
         \midrule
         Gaussian &346 & 0.032\\
         \midrule
         Gaussian &3304 & 0.026\\
         Box &3304 & 0.026\\
         \bottomrule
    \end{tabular}
    \caption{The $3\sigma$ discovery potential fluences for different data sets, time pdfs (first row), and data sets (second row). The second row lists the duration of the IceCube data taking period within the respective data set. We compare the fluences for a Gaussian time pdf with the fluence for a box-shaped time pdf for a flare in the latest data set (last two lines). We find that the fluence does not depend on the shape of the time pdf.}
    \label{tab:three-sigma-discovery-potential}
\end{table}




\section{Conclusion}
We search for transient neutrino sources in 12 years of IceCube archival data. We present improved methods for a time-dependent point source analysis. With expectation maximization we use a fast and efficient flare finding method. We also explore methods to analytically describe how the test statistic quantiles change for different flare intensities. We determine $3\sigma$ discovery potential fluences in the range of $\approx 2.7 \cdot 10^{-2} \rm{GeV}/\rm{cm}^2$. With this analysis, we will be able set new upper fluence limits as soon as we run the analysis on unblinded IceCube data.

\bibliographystyle{ICRC}
\bibliography{references}



\clearpage
\section*{Full Author List: IceCube Collaboration}




\scriptsize
\noindent
R. Abbasi$^{17}$,
M. Ackermann$^{59}$,
J. Adams$^{18}$,
J. A. Aguilar$^{12}$,
M. Ahlers$^{22}$,
M. Ahrens$^{50}$,
C. Alispach$^{28}$,
A. A. Alves Jr.$^{31}$,
N. M. Amin$^{42}$,
R. An$^{14}$
% ...
\\

\noindent
% ...
$^{12}$ Universit{\'e} Libre de Bruxelles, Science Faculty CP230, B-1050 Brussels, Belgium \\
% ...
$^{14}$ Department of Physics and Laboratory for Particle Physics and Cosmology, Harvard University, Cambridge, MA 02138, USA \\
% ...
$^{17}$ Department of Physics, Loyola University Chicago, Chicago, IL 60660, USA \\
$^{18}$ Dept. of Physics and Astronomy, University of Canterbury, Private Bag 4800, Christchurch, New Zealand \\
% ...
$^{22}$ Niels Bohr Institute, University of Copenhagen, DK-2100 Copenhagen, Denmark \\
% ...
$^{28}$ D{\'e}partement de physique nucl{\'e}aire et corpusculaire, Universit{\'e} de Gen{\`e}ve, CH-1211 Gen{\`e}ve, Switzerland \\
% ...
$^{31}$ Karlsruhe Institute of Technology, Institute for Astroparticle Physics, D-76021 Karlsruhe, Germany  \\
% ...
$^{42}$ Bartol Research Institute and Dept. of Physics and Astronomy, University of Delaware, Newark, DE 19716, USA \\
% ...
$^{50}$ Oskar Klein Centre and Dept. of Physics, Stockholm University, SE-10691 Stockholm, Sweden \\
% ...
$^{59}$ DESY, D-15738 Zeuthen, Germany \\
% ...

\subsection*{Acknowledgements}

\noindent
USA {\textendash} U.S. National Science Foundation-Office of Polar Programs,
U.S. National Science Foundation-Physics Division,
U.S. National Science Foundation-EPSCoR,
Wisconsin Alumni Research Foundation,
Center for High Throughput Computing (CHTC) at the University of Wisconsin{\textendash}Madison,
Open Science Grid (OSG),
Extreme Science and Engineering Discovery Environment (XSEDE),
Frontera computing project at the Texas Advanced Computing Center,
U.S. Department of Energy-National Energy Research Scientific Computing Center,
Particle astrophysics research computing center at the University of Maryland,
Institute for Cyber-Enabled Research at Michigan State University,
and Astroparticle physics computational facility at Marquette University;
Belgium {\textendash} Funds for Scientific Research (FRS-FNRS and FWO),
FWO Odysseus and Big Science programmes,
and Belgian Federal Science Policy Office (Belspo);
Germany {\textendash} Bundesministerium f{\"u}r Bildung und Forschung (BMBF),
Deutsche Forschungsgemeinschaft (DFG),
Helmholtz Alliance for Astroparticle Physics (HAP),
Initiative and Networking Fund of the Helmholtz Association,
Deutsches Elektronen Synchrotron (DESY),
and High Performance Computing cluster of the RWTH Aachen;
Sweden {\textendash} Swedish Research Council,
Swedish Polar Research Secretariat,
Swedish National Infrastructure for Computing (SNIC),
and Knut and Alice Wallenberg Foundation;
Australia {\textendash} Australian Research Council;
Canada {\textendash} Natural Sciences and Engineering Research Council of Canada,
Calcul Qu{\'e}bec, Compute Ontario, Canada Foundation for Innovation, WestGrid, and Compute Canada;
Denmark {\textendash} Villum Fonden and Carlsberg Foundation;
New Zealand {\textendash} Marsden Fund;
Japan {\textendash} Japan Society for Promotion of Science (JSPS)
and Institute for Global Prominent Research (IGPR) of Chiba University;
Korea {\textendash} National Research Foundation of Korea (NRF);
Switzerland {\textendash} Swiss National Science Foundation (SNSF);
United Kingdom {\textendash} Department of Physics, University of Oxford.

\end{document}